\documentclass[twocolumn,superscriptaddress,
showpacs,preprintnumbers,amsmath,amssymb]{revtex4}

\usepackage{amssymb,amsmath,amsthm}
\usepackage{epsfig}
\usepackage{multirow}
\newtheorem{Thm}{Theorem}

\newtheorem{Lem}[Thm]{Lemma}

\theoremstyle{definition}

\newcommand{\bra}[1]{{\left\langle #1 \right|}}
\newcommand{\ket}[1]{{\left| #1 \right\rangle}}
\newcommand{\inn}[2]{{\left\langle #1 | #2 \right\rangle}}

\newcommand{\T}{\mbox{$\mathrm{tr}$}}
\newcommand{\K}{\mbox{$\mathcal H$}}
\def\a{\alpha}
\def\b{\beta}
\def\c{\gamma}
\def\d{\delta}

\begin{document}
\title{Non-static Quantum Bit Commitment}

\author{Jeong Woon Choi}\email{jw_choi@etri.re.kr}
\affiliation{
 Information Security Research Division,
Electronics and Telecommunications Research Institute, Daejeon
305-700, Korea }
\author{Dowon Hong}
\affiliation{
 Information Security Research Division,
Electronics and Telecommunications Research Institute, Daejeon
305-700, Korea }
\author{Ku-Young Chang}
\affiliation{
 Information Security Research Division,
Electronics and Telecommunications Research Institute, Daejeon
305-700, Korea }
\author{Dong Pyo Chi}
\affiliation{
 Department of Mathematical Sciences,
 Seoul National University, Seoul 151-747, Korea}
\author{Soojoon Lee}
\affiliation{ Department of Mathematics and Research Institute for
Basic Sciences,
Kyung Hee University, Seoul 130-701, Korea}

\date{\today}

\begin{abstract}
Quantum bit commitment has been known to be impossible by the
independent proofs of Mayers, and Lo and Chau, under the
assumption that the whole quantum states right before the
unveiling phase are static to users. We here provide an
unconditionally secure non-static quantum bit commitment protocol with a trusted third party,
which is not directly involved in any communications between users and
can be limited not to get any information of commitment without being detected by users.
We also prove that our quantum bit commitment protocol is not secure without the help of the trusted
third party. The proof is basically different from the Mayers-Lo-Chau's no-go
theorem, because we do not assume the staticity of the finally
shared quantum states between users.

\end{abstract}

\pacs{ 
03.67.Dd, 
03.67.Hk, 
03.67.Mn  
}
\maketitle

\section{Introduction}
As one of the most basic and important cryptographic primitives, a bit commitment (BC) scheme
has a lot of applications to crucial cryptographic
protocols including coin flipping, interactive zero-knowledge
proof, oblivious transfer, verifiable secret sharing, multiparty
secure computation, and so on~\cite{B,BCC,GMR,GMW,BBCS,K01}.
There have also been several quantum approaches~\cite{BC,BCJL} to guarantee the unconditional security of BC protocols,
as quantum key distribution (QKD) protocols~\cite{BB,LC02} have done.
Unfortunately, in the middle of the 1990's Mayers~\cite{M01,M02}, and
independently Lo and Chau~\cite{LC} (MLC) proved that quantum principles
cannot be helpful to construct an unconditionally secure BC protocol, in contrast to a brilliant development of QKD protocols~\cite{H01,U,SRS}.
The impossibility of quantum bit commitment (QBC) is called the MLC's no-go theorem, which
implies a severe drawback of quantum cryptography.
Since then, there have been several results about QBC protocols, some of which are for the possibility through new schemes and theories~\cite{K,HK,H02},
others of which are for the trade-off relations between the possibility and the impossibility~\cite{SR,DKSW}.

The most important assumption of the MLC's no-go theorem
is that every QBC protocol results in a \textit{static} quantum state, and thus
both users exactly know about what it is before the unveiling time.
For any initial states of Alice and Bob, $\ket{\chi}_A$ ($\chi=0$ or $1$ ) and $\ket{\psi}_B$, the finally shared quantum state will be given as
$U_{AB}(\ket{\chi}_A\otimes\ket{\psi}_B)$, where $U_{AB}$ represents all the algorithms involved in the protocol
and is necessarily opened and known to all participants.
If the QBC protocol satisfies the perfect concealment, then by the Gisin-Hughston-Jozsa-Wootters (GHJW) theorem~\cite{GHJW}
there exists a local unitary operation $S_A$
such that $(S_A\otimes I)U_{AB}(\ket{0}_A\otimes\ket{\psi}_B)=U_{AB}(\ket{1}_A\otimes\ket{\psi}_B)$.
By delaying the measurements and applying $S_A$ to the local system, Alice is able to change her committed bit surreptitiously without being detected by Bob.
This is the main stream of the MLC's no-go theorem.

However, we focus on the fact that $S_A$ actually is given depending on the Bob's initial state $\ket{\psi}_B$.
So, it would be better to denote the Alice's strategy by $S_A(\psi)$ rather than $S_A$.
Even though it is true that there exists an exact operation $S_A(\psi)$ for each $\ket{\psi}_B$ whenever the protocol is perfectly concealing,
Alice could neither figure out nor make use of $S_A(\psi)$ appropriately, if $\ket{\psi}_B$ is randomly given and kept unknown to her.
A QBC protocol to realize the above situation is here called a \textit{non-static} QBC protocol.

In this paper, by investigating the possibility and the impossibility of such non-static QBC protocols,
we construct an unconditionally secure QBC protocol with the help of a trusted third party (TTP), and prove that our non-static QBC protocol is not possible without the help of a TTP.
Although the existence of a TTP can be a weak point as in general cryptographic primitives,
the TTP in our protocol plays only a little role to provide quantum sources to carry classical bit information.
Moreover, the TTP is not actually involved in any communications between users, and
cannot get any information about the commitment without being detected by users.

\section{Non-static QBC Protocols }
Hereafter we consider a more generalized version of QBC protocols
which varies the resulting states according to the initial state $\ket{\psi}_B$ generated by Bob (or a TTP), and thus
the strategy $S_A(\psi)\otimes I$ by a dishonest Alice might be also changed according to $\ket{\psi}_B$.
One possible way to accomplish the above property is that Bob (or a TTP), instead of Alice, prepares and sends an initial quantum state $\ket{\psi}_B$ to Alice,
where $\ket{\psi}_B$ should be kept unknown to Alice.
Then Alice applies an associate unitary operator to  $\ket{\psi}_B$ to commit a bit $\chi$.

For example, suppose that when $\chi=0$, Alice chooses one of $M$ and $N$ randomly, and similarly
when $\chi=1$, one of $J$ and $K$ randomly, where $M$, $N$, $J$, and $K$ are defined as
\begin{eqnarray}
M &=& I,\quad N=-i\sigma_y \nonumber\\
J &=& \frac{1-i}{2\sqrt{2}} \left[ I + i\left(\sigma_x-\sigma_y+\sigma_z\right)\right]\nonumber\\
K &=& \frac{1+i}{2\sqrt{2}} \left[ I + i\left(\sigma_x+\sigma_y-\sigma_z\right)\right],
\end{eqnarray}
where $\sigma_x$, $\sigma_y$, and $\sigma_z$ are the Pauli matrices.
To guarantee the randomness, Alice prepares an auxiliary state
$\ket{+}_A=\frac{\ket{0}_A+\ket{1}_A}{\sqrt{2}}$, and then she applies
a unitary operator either $\ket{0}_A\bra{0}\otimes M +
\ket{1}_A\bra{1}\otimes N$ (if $\chi=0$) or $\ket{0}_A\bra{0}\otimes
J + \ket{1}_A\bra{1}\otimes K $ (if $\chi=1$) to $\ket{+}_A\otimes\ket{\psi}_B$ so that she finally obtains the following states
\begin{eqnarray}
\ket{\Phi_0(\psi)}_{AB}&=&\dfrac{\ket{0}_{A}\otimes M \ket{\psi}_{B}+\ket{1}_{A}\otimes N \ket{\psi}_{B}}{\sqrt{2}} ~\text{ and}\nonumber\\
\ket{\Phi_1(\psi)}_{AB}&=&\dfrac{\ket{0}_{A}\otimes J \ket{\psi}_{B}+\ket{1}_{A}\otimes K \ket{\psi}_{B}}{\sqrt{2}}.
\end{eqnarray}
By performing the standard measurement on her local system $\K_A$, Alice provides
Bob with an uniformly distributed ensemble, either
$\xi_0(\psi)=\{M\ket{\psi}_B,N\ket{\psi}_B\}$ or
$\xi_1(\psi)=\{J\ket{\psi}_B,K\ket{\psi}_B\}$
as shown in TABLE~I.
\begin{table}[h]\label{1}
\caption{The change of the initial states $\ket{\psi}_B=m\ket{0}_B+n\ket{1}_B$ ($|m|^2+|n|^2=1$): It
shows how the initial states $\ket{\psi}_B$ are transformed by unitary operators $M,N,J$, and $K$ randomly chosen according to $\chi$.
($\ket{\pm}_B$ denotes $\frac{\ket{0}_B\pm\ket{1}_B}{\sqrt{2}}$.) }
\begin{center}
\begin{tabular}{c|c|c|c|c}
\hline
  \hline
  $\chi$& Operators & $\ket{0}_B$ & $\ket{1}_B$ & $\ket{\psi}_B=m\ket{0}_B+n\ket{1}_B$   \\
  \hline
  \hline
  \multirow{2}{0.2cm}[0cm]{$0$} & $M$ & $\ket{0}_B$ & $\ket{1}_B$ & $m\ket{0}_B+n\ket{1}_B$  \\
  \cline{2-5}
   & $N$ & $\ket{1}_B$ & $-\ket{0}_B$ & $m\ket{1}_B-n\ket{0}_B$  \\
  \hline
  \multirow{2}{0.2cm}[0cm]{$1$} & $J$ & $\ket{+}_B$ & $i\ket{-}_B$ & $m\ket{+}_B+in\ket{-}_B$ \\
  \cline{2-5}
   & $K$ & $\ket{-}_B$ & $i\ket{+}_B$ & $m\ket{-}_B+in\ket{+}_B$  \\
  \hline
  \hline
\end{tabular}
\end{center}
\end{table}

Without an additional information about the ensembles, Bob will
regard them as a density operator, either
$\rho_0(\psi)=(M\ket{\psi}_B\bra{\psi}M^{\dagger}+N\ket{\psi}_B\bra{\psi}N^{\dagger})/2$
or
$\rho_1(\psi)=(J\ket{\psi}_B\bra{\psi}J^{\dagger}+K\ket{\psi}_B\bra{\psi}K^{\dagger})/2$,
respectively.

Let us consider the cases that $\ket{\psi}_B=\ket{0}_B$ and $\ket{\psi}_B=\ket{+}_B$.
It is very easy to show that $\rho_0(\psi=0)=\rho_1(\psi=0)=\rho_0(\psi=+)=\rho_1(\psi=+)=I/2$. However, we can ask a question such as ``Is there
any proper strategy $S_A$ to change not only $\ket{\Phi_0(\psi=0)}_{AB}$ to $\ket{\Phi_1(\psi=0)}_{AB}$ but also $\ket{\Phi_0(\psi=+)}_{AB}$ to $\ket{\Phi_1(\psi=+)}_{AB}$?"
The answer is NO. In fact, up to the left multiplication of diagonal matrices,
$S_A(\psi=0)$ should be $\dfrac{1}{\sqrt{2}}\begin{pmatrix}
  1 & 1 \\
  1 & -1 \\
\end{pmatrix}$, while
$S_A(\psi=+)$ should be $\dfrac{1}{\sqrt{2}}\begin{pmatrix}
  1 & -i \\
  i & -1 \\
\end{pmatrix}$. This means that a certain fixed attack by Alice cannot be available for all $\ket{\psi}_B$, and therefore Alice should
be able to choose a strategy appropriate to an unknown $\ket{\psi}_B$.

However, this example has a problem that the QBC protocol is not perfectly concealing.
If Bob prepares the initial state as $\ket{\psi}_B=\frac{\ket{0}_B+i\ket{1}_B}{\sqrt{2}}$, then
he can know Alice's commitment in advance, because $\rho_0$ and $\rho_1$ are obviously different.
To solve this problem, we employ a TTP, and then investigate
the securities of QBC protocols with and without the help of the TTP in the next two subsections.

\subsection{Non-static QBC Protocol with a TTP}
Alice and TTP previously share $\mathcal{N}$ maximally entangled states $\ket{\Psi^-}_{TA}=(\ket{01}_{TA}-\ket{10}_{TA})/\sqrt{2}$
satisfying $\ket{\Psi^-}_{TA}=(U\otimes U)\ket{\Psi^-}_{TA}$
up to the global phase for all unitary operators $U$.

(i) [Pre-Commitment] TTP performs random orthogonal measurements
$M_i=\{\ket{\phi_i}_T\bra{\phi_i}_T,\ket{\phi^{\perp}_i}_T\bra{\phi^{\perp}_i}_T \}$ ($1 \leq i \leq \mathcal{N}$) on his side of $\ket{\Psi^-}_{TA}$'s.
Then Alice and TTP always have the opposite state, that is, if TTP's result is $\ket{\phi_i}_T$ ($\ket{\phi^{\perp}_i}_T$), then Alice must have $\ket{\psi_i}_A=\ket{\phi^{\perp}_i}_A$ ($\ket{\phi_i}_A$).
However, Alice does not know what $\ket{\psi_i}_A$'s are actually, because TTP keeps $M_i$ unknown to her.

(ii) [Commitment] To commit a bit $\chi$, Alice encodes $\chi$ into $\ket{\psi_i}_A$
by applying an operator $P_i$ randomly chosen from $M$, $N$, $J$, and $K$ as follows.
If Alice wants to commit $0$, then she sends Bob
$M \ket{\psi_i}_A$ or $N \ket{\psi_i}_A$ at random, and if she wants to commit $1$, then she sends $J\ket{\psi_i}_A$ and $K \ket{\psi_i}_A$ randomly.

(iii) [Holding Phase] It proceeds without doing anything for a certain period which users agreed with at the beginning stage of
the protocol.

(iv) [Unveiling Phase] At a specific later time, Alice publicly announces all $P_i$'s and then TTP all $M_i$'s and measurement outcomes.
Then Bob verifies the commitment by checking whether the measurement outcomes are always opposite or not, when he performs $M_i$'s on $P_i^{\dagger}P_i\ket{\psi_i}_A$.
If Alice is honest, then the measurement outcomes should be opposite for all $i$.

In step (ii), as noticed previously, by using the ancillary state $\ket{+}_{A'}$ and the non-local unitary operations such as
$\ket{0}_A\bra{0}\otimes M +\ket{1}_A\bra{1}\otimes N$ and $\ket{0}_A\bra{0}\otimes J +\ket{1}_A\bra{1}\otimes K$ according to $\chi$,
Alice obtains $\ket{\Phi_0}_{A'A} = \left(\ket{0}_{A'} \otimes M\ket{\psi}_A + \ket{1}_{A'} \otimes N\ket{\psi}_A \right)/\sqrt{2}$ and
$\ket{\Phi_1}_{A'A} = \left(\ket{0}_{A'} \otimes J\ket{\psi}_A + \ket{1}_{A'} \otimes K\ket{\psi}_A \right)/\sqrt{2}.$
However, due to the randomness of $\ket{\psi}_A$, $\ket{\Phi_\chi}_{A'A}$ will be changed every time. These states can come to not only product states but also
maximally entangled state. So, Alice could not control the relation between $\ket{\Phi_0}_{A'A}$ and $\ket{\Phi_1}_{A'A}$ as she wants, without the knowledge of $\ket{\psi}_A$'s
(actually $M_i$'s).

Of course, we need to calculate the success probability of the delayed measurement attack proposed in the MLC's no-go theorem,
which can be measured with the fidelity $F(\ket{\psi},\ket{\phi})=|\inn{\psi}{\phi}|^2$.
Suppose that, to change the committed bit from $0$ to $1$, Alice applies a local unitary operation $\begin{pmatrix}
  a & b \\
  c & d \\
\end{pmatrix}$.
The success probability is
\begin{eqnarray}
\mathbb{F}=\frac{1}{2} \{F(aM\ket{\psi}_A+bN\ket{\psi}_A,J\ket{\phi}_A) \nonumber\\
+F(cM\ket{\psi}_A+dN\ket{\psi}_A,K\ket{\phi}_A)\},
\end{eqnarray}
and therefore, in the Bloch representation, $\ket{\psi}_A=\cos(\theta/2)\ket{0}_A+e^{i\phi}\sin(\theta/2)\ket{1}_A$ ($0 \leq \theta \leq \pi, 0 \leq \phi \leq 2\pi$),
the expected success probability is
\begin{eqnarray}
&&\frac{1}{4\pi}\int_0^{2\pi}\int_0^{\pi}\mathbb{F} \sin\theta ~d\theta d\phi \nonumber\\
\quad &=& \frac{|a|^2+|b|^2+|c|^2+|d|^2}{4}+\frac{Re(a\overline{b}-c\overline{d})}{6} \nonumber\\
\quad &=& \frac{1}{2}+\frac{2Re(a\overline{b})}{6}
\leq \frac{1}{2}+\frac{|a\overline{b}|}{3} \leq \frac{2}{3},
\end{eqnarray}
where $\overline{z}$ is the complex conjugate of a given complex number $z$.
Since the protocol is repeated $\mathcal{N}$ times, Alice's attack is detected with the probability greater than $1-(2/3)^\mathcal{N}$ which goes to $1$
as $\mathcal{N} \rightarrow \infty$.
That is, this QBC protocol satisfies the asymptotic bindingness, where the level of security follows as noticed in~\cite{SR}.

We should also consider the concealment.
One of the assumptions of our protocol is that Alice and TTP previously share the singlet states. This means that
Bob has no way to interrupt the quantum channel between them to get some information. That is to say, Bob should gain information about the commitment
from only quantum states given by Alice. Another assumption is that TTP should choose $M_i$'s at true random. So, the finally encoded states
will appear to Bob as $I/2$, which guarantees the perfect concealment.

To transmit only digital information through classical channels, TTP can choose the bases of $M_i$'s in a discretized subset of the Bloch space.
For instance, TTP can select finite points uniformly dividing the sub-circle spanned by $\ket{0}$, $\ket{1}$, $\ket{+}$ and $\ket{-}$.
Since our protocol satisfies the perfect concealment for all initial states $\ket{\psi}_A$ such that $m\overline{n} \in \mathbb{R}$,
so do all points in the sub-circle. Of course, the success probability will be changed a little bit but less than 1, and therefore this protocol still satisfies the bindingness.
Such a restriction on the domain of initial states gives us one more advantage, which prohibits
TTP from generating the initial states such that $m\overline{n} \notin \mathbb{R}$ and knowing Alice's commitment in advance.
TTP should always announce Bob the right information about his measurements, because if TTP announces dishonestly, then the measurements in the wrong bases
will make a disturbance on the correlation between Alice and Bob, and thus the dishonest behavior will be detected by users.

In result, the quantum entanglement shared between Alice and TTP guarantees not only the non-staticity, but therefore also the unconditional security of our protocol,
which cannot be realized by the classical cryptographic theories.

\subsection{Non-static QBC Protocol without a TTP}

We here deal with a self-enforcing QBC protocol (without a TTP),
which is slightly modified from our previous QBC protocol like that Bob, instead of TTP, generates initial quantum states $\ket{\psi}_B$ and Alice applies unitary operators to $\ket{\psi}_B$ to commit $\chi$.
%
%

The following lemma is a necessary and sufficient condition for our self-enforcing QBC protocol to be perfectly concealing
against Bob using any kind of quantum entangled state $\ket{\Psi}_{BB'}$ on the extended system $\K_{B}\otimes \K_{B'}$.
\begin{Lem}\label{Concealment}
A non-static QBC protocol is perfectly concealing for all qubits $\ket{\psi}_B$ and all entangled state $\ket{\Psi}_{BB'}$
if and only if $M$, $N$, $J$, and $K$ should satisfy the following equations
\begin{eqnarray}
M\ket{0}_B\bra{0}M^{\dagger}+N\ket{0}_B\bra{0}N^{\dagger}&=&J\ket{0}_B\bra{0}J^{\dagger}+K\ket{0}_B\bra{0}K^{\dagger},\nonumber\\
M\ket{1}_B\bra{1}M^{\dagger}+N\ket{1}_B\bra{1}N^{\dagger}&=&J\ket{1}_B\bra{1}J^{\dagger}+K\ket{1}_B\bra{1}K^{\dagger},\nonumber\\
&and&\\
M\ket{0}_B\bra{1}M^{\dagger}+N\ket{0}_B\bra{1}N^{\dagger}&=&J\ket{0}_B\bra{1}J^{\dagger}+K\ket{0}_B\bra{1}K^{\dagger}.\nonumber
\end{eqnarray}
\end{Lem}
\begin{proof}
By a direct calculation, we first prove that the above condition is a necessary and sufficient condition for $\rho_0(\psi)=\rho_1(\psi)$ for all qubits $\ket{\psi}_B=m\ket{0}_B+n\ket{1}_B$.
It is very clear that if $M$, $N$, $J$, and $K$ satisfy Eq.~(5), then $\rho_0(\psi)=\rho_1(\psi)$.
Conversely, we should show that all $M$, $N$, $J$, and $K$ such that $\rho_0(\psi)=\rho_1(\psi)$ satisfy Eq.~(5).
The first two equations of Eq.~(5) can be easily derived from the cases that $m\neq 0, n=0$ and $m=0, n\neq 0$.
Therefore, $M$, $N$, $J$, and $K$ should eventually satisfy
$\mathrm{Re}\left(m\overline{n}(M\ket{0}_B\bra{1}M^{\dagger}+N\ket{0}_B\bra{1}N^{\dagger})\right) = \mathrm{Re}\left(m\overline{n}(J\ket{0}_B\bra{1}J^{\dagger}+K\ket{0}_B\bra{1}K^{\dagger})\right)$,
for all $m$ and $n$.
Considering the cases that $m=n=1$ and $m=1, n=i$, we can obtain the third equation of Eq.~(5).
It is trivial to extend the necessary and sufficient condition to all bipartite entangled states $\ket{\Psi}_{BB'}$ on $\K_B \otimes \K_{B'}$, where $\dim \K_B=2$ and $\dim \K_{B'}$ is arbitrary,
because $\ket{\Psi}_{BB'}$ has the Schmidt decomposition~\cite{SD} and we can regard $\ket{0}_B$ and $\ket{1}_B$ as eigenvectors of the density operator $\T_{B'}(\ket{\Psi}_{BB'}\bra{\Psi})$.
\end{proof}

In addition, we also figure out what kind of unitary operators $M$, $N$, $J$, and $K$
are able to satisfy the perfect concealment, that is, the necessary and sufficient condition given in Lemma~\ref{Concealment}.
Unfortunately, Theorem~\ref{localattack} tells us that there is a strategy for Alice to cheat the commitment freely,
regardless of whether she knows the initial quantum states or not.
\begin{Thm}~\label{localattack}
If a non-static QBC protocol is perfectly concealing, then there exists a local unitary operator $S_{A}=\left(%
\begin{array}{cc}
  a & b \\
  c & d \\
\end{array}%
\right)$ such that $J=aM+bN$ and $K=cM+dN$.
\end{Thm}
\begin{proof}
Considering the orthogonality and the GHJW theorem for the perfect concealment, we can let $M$, $N$, $J$, and $K$ be unitary matrices as shown in TABLE~II, without loss of generality.
\begin{table}[h]\label{2}
\caption{The parametrization for the unitary operators $M,N,J$, and $K$ satisfying that $\rho_0(\psi)=\rho_1(\psi)$ for all $\ket{\psi}_B$:}
$|x|^2+|y|^2=1$ ($y \neq 0$), $|\a|=1$, $\begin{pmatrix}
  a & b \\
  c & d \\
\end{pmatrix},\begin{pmatrix}
  s & t \\
  u & v \\
\end{pmatrix}$ : unitary

\begin{center}
\begin{tabular}{c|c|c}
\hline
  \hline
  Operators & $\ket{0}_B$ & $\ket{1}_B$   \\
  \hline
  \hline
  $M$ & $\ket{0}_B$ & $\ket{1}_B$   \\
  $N$ & $x\ket{0}_B+y\ket{1}_B$ & $\a(\overline{y}\ket{0}_B-\overline{x}\ket{1}_B)$   \\
  \hline
  $J$ & $(a+bx)\ket{0}_B+by\ket{1}_B$ & $t\a \overline{y}\ket{0}_B+(s-t\a \overline{x})\ket{1}_B$  \\
  $K$ & $(c+dx)\ket{0}_B+dy\ket{1}_B$ & $v\a \overline{y}\ket{0}_B+(u-v\a \overline{x})\ket{1}_B$  \\
  \hline
  \hline
\end{tabular}
\end{center}
\end{table}
In this case, it is obvious that $M$, $N$, $J$, and $K$ satisfy the first two equations of Eq.~(5).
By the third equation of Eq.~(5), all parameters in TABLE II should follow that
\begin{eqnarray}
\overline{\a} x y &=& (a+bx)\overline{t\a} y + (c+dx)\overline{v\a} y, \nonumber\\
-y\overline{\a} x &=& by(\overline{s}-\overline{t\a} x)+dy(\overline{u}-\overline{v\a} x),\nonumber\\
1-\overline{\a} x^2 &=& (a+bx)(\overline{s}-\overline{t\a} x) +(c+dx)(\overline{u}-\overline{v\a} x),\nonumber\\
\overline{\a} y^2 &=& by^2 \overline{t\a} + dy^2\overline{v\a}.
\end{eqnarray}
We first consider the case that $\rho_0(\psi=0)$ is invertible (of rank 2), that is, $y\neq 0$. Eq.~(6) can be rewritten as
\begin{eqnarray}
1 &=& a\overline{s}+c\overline{u}, \nonumber\\
0 &=& b\overline{s}+d\overline{u}, \nonumber\\
0 &=& a\overline{t}+c\overline{v}, \text{ and }\nonumber\\
1 &=& b\overline{t}+d\overline{v}.
\end{eqnarray}
This means that $\begin{pmatrix}
  \overline{s} & \overline{t} \\
  \overline{u} & \overline{v} \\
\end{pmatrix}\begin{pmatrix}
  a & b \\
  c & d \\
\end{pmatrix}=\begin{pmatrix}
  1 & 0 \\
  0 & 1 \\
\end{pmatrix}$, that is, $\begin{pmatrix}
  a & b \\
  c & d \\
\end{pmatrix}=\begin{pmatrix}
  s & t \\
  u & v \\
\end{pmatrix}$. Therefore, there exists a unitary operator $\begin{pmatrix}
  a & b \\
  c & d \\
\end{pmatrix}$ such that $J=aM+bN$ and $K=cM+dN$.

Let us consider the case that the rank of $\rho_0(\psi=0)$ is 1, that is $y=0$, where we can reparameterize $M$, $N$, $J$, and $K$ as shown in TABLE~III.
\begin{table}[h]\label{3}
\caption{The reparametrization of TABLE II for the unitary operators $M,N,J$, and $K$ satisfying that $\mathrm{rank}(\rho_0(\psi=0))=1$:}
$|j|=|k|=|l|=|m|=1$, $|\a|=|\b|=|\c|=|\d|=1$
\begin{center}
\begin{tabular}{c|c|c}
\hline
  \hline
  Operators & $\ket{0}_B$ & $\ket{1}_B$   \\
  \hline
  \hline
  $M$ & $j\ket{0}_B$ & $k\ket{1}_B$   \\
  $N$ & $l\ket{0}_B$ & $m\ket{1}_B$   \\
  \hline
  $J$ & $\a\ket{0}_B$ & $\b\ket{1}_B$   \\
  $K$ & $\c\ket{0}_B$ & $\d\ket{1}_B$   \\
  \hline
  \hline
\end{tabular}
\end{center}
\end{table}
For the perfect concealment, the parameters should satisfy
\begin{eqnarray}
j\overline{k}+l\overline{m}=\a\overline{\b}+\c\overline{\d}.
\end{eqnarray}
If $j\overline{k}+l\overline{m} \neq 0$, then $j\overline{k}=\a\overline{\b}$, $l\overline{m}=\c\overline{\d}$
or $j\overline{k}=\c\overline{\d}$, $l\overline{m}=\a\overline{\b}$, because of the unity of parameters.
This property means that the matrices have the relations such as $M\propto J$, $N\propto K$ or $M\propto K$, $N\propto J$, where
$A\propto B$ denotes $A=cB$ for a constant $c$.
Therefore, the commitments according to $\chi$'s are actually
same and thus make no sense.
If $j\overline{k}+l\overline{m}=0$, under the assumption that $j\overline{k}\neq l\overline{m}$
(Otherwise, for all quantum states $\ket{\psi}_B$, $\mathrm{rank}(\rho_0(\psi))=\mathrm{rank}(\rho_1(\psi))=1$, and thus $M\propto N\propto J\propto K$, which is meaningless.),
we can find a unitary operator $\begin{pmatrix}
  a & b \\
  c & d \\
\end{pmatrix}$ such that $J=aM+bN$ and $K=cM+dN$, where $a$, $b$, $c$, and $d$ are given as
\begin{eqnarray*}
a&=&\frac{l\b-m\a}{lk-mj}, b=\frac{k\a-j\b}{lk-mj}, c=\frac{l\d-m\c}{lk-mj}, \text{ and }\\
d&=&\frac{k\c-j\d}{lk-mj}.
\end{eqnarray*}
This completes the proof.
\end{proof}

By using $S_{A}\otimes I$, Alice can freely exchange unitary operators $M$ and $N$ with $J$ and $K$ so that she can cheat her committed bit with certainty
without being detected by Bob. Therefore, we can find out that, even though dishonest Bob makes use of arbitrary dimensional ancillary system,
if Alice and Bob communicate through the only two-dimensional channel,
then any non-static QBC protocols we propose are not secure, and in fact, the perfect concealment makes the non-static QBC protocol static without the help of a TTP.


\section{Conclusion}\label{Sec:conclusion}
We have dealt with a new QBC scheme which can be not static so that
the final quantum states are determined randomly and kept unknown to all participants until the unveiling phase.
However, we would like to emphasize that our QBC scheme does not oppose the MLC's no-go theorem, but ensures its security only by enforcing Alice
to change the attack strategy according to the unknown initial quantum information.

We have shown that it is possible to construct an unconditionally secure QBC protocol with the help of a TTP, where
the role of the TTP can be limited not to get any information of the committed bit in advance and actually users can perceive any dishonest behaviors of the TTP.
Unfortunately, we have also proved that the non-static QBC protocol is not secure without the help of the TTP.
In a self-enforcing non-static QBC protocol, the necessary and sufficient condition for the perfect concealment eventually
makes the QBC protocol static.
It would be important to check if we can extend the impossibility of the self-enforcing QBC protocols to the cases with no limits on
the dimension of quantum channels and the number of the quantum states in ensembles.

\section*{Acknowledgments}
This work was supported by the IT R\&D program of MKE/IITA (Grant
No. 2005-Y-001-05, ``Developments of next generation security
technology'' and Grant No. 2008-F-035-02, ``Development of Key
Technologies for Commercial Quantum Cryptography Communication
System''). D.P.C. was supported by a Korea Science and Engineering
Foundation (KOSEF) grant funded by the Korean Government (MOST).
S.L. was supported by Basic Science Research Program
through the National Research Foundation of Korea (NRF)
funded by the Ministry of Education, Science and Technology (Grant No. 2009-0076578).


\end{document}